TMUP-HEL-9508
October, 1995

# Extended Technicolor Contribution to the $Zbb$ Vertex *)

Noriaki KITAZAWA **)

*Department of Physics, Tokyo Metropolitan University,*
*Hachioji-shi, Tokyo 192-03, Japan*

**Abstract**

The deviation of the experimental value of $R_b \equiv \Gamma_b/\Gamma_h$ from the standard-model prediction can be naturally explained in the extended technicolor scenario. The diagonal extended technicolor gauge boson which flavor-diagonally couples with both the ordinary fermions and the techni-fermions plays an important role in order to have the appropriate radiative correction to the $Zbb$ vertex. But since the diagonal extended technicolor gauge boson gives too large positive contribution to the $T$ parameter, we need some new mechanisms which generate negative contribution to the $T$ parameter.

*) Based on the talk given at Yukawa International Seminar '95: From the Standard Model to Grand Unified Theories, Kyoto, Japan, Aug. 21-25.

**) E-mail address: kitazawa@musashi.phys.metro-u.ac.jp

## §1. Introduction

The measurement of the quantity $R_b \equiv \Gamma_b/\Gamma_h$ at LEP shows a large deviation from the prediction of the standard model. The measured value $R_b = 0.2202 \pm 0.0020$ deviates at 2-$\sigma$ level from the standard-model prediction $R_b = 0.2157$ ($m_t = 175$ GeV)[1,2]. This may be the signature of the new physics on $Zbb$ vertex.

It has been pointed out that the "sideways" gauge boson of the extended technicolor (ETC) theory generates significant correction to the $Zbb$ vertex[3]. The reason is that the relatively light ($\mathcal{O}(1)$ TeV) sideways boson associated with the top quark mass generation should couple with the left-handed bottom quark according to the $SU(2)_L$ symmetry. This contribution is highly model independent. The "diagonal" gauge bosons which appears in the most ETC models also generates the correction to the $Zbb$ vertex[4]. The magnitude of this diagonal contribution is comparable with the sideways contribution and the sign is opposite[5]. The sideways and the standard-model contributions make $R_b$ small, while the diagonal contribution makes it large. Therefore, if the diagonal contribution is large enough to cancel out the other two contributions, the LEP result can be explained.

In section 2, we show that the diagonal contribution can naturally explain the LEP result of $R_b$ in some models of the ETC theory[6]. We also point out that in the presence of such a large correction to the $Zbb$ vertex the value of the QCD gauge coupling $\alpha_s(m_Z)$ which is extracted from the $Z$ boson data becomes more consistent with both the recent Lattice-QCD evaluation[7] and the global average of the Particle Data Group[8][6].

In section 3, we estimate the contribution of the diagonal ETC boson to the $T$ parameter[9] assuming that the value of $R_b$ is explained by the diagonal ETC boson. As already pointed out by Yoshikawa[10], it is shown that the contribution is too large beyond the experimental bound. We conclude that some mechanisms which generate negative contribution to the $T$ parameter are needed to explain the the experimental value of $R_b$ by the diagonal ETC boson.

## §2. Diagonal ETC boson and $Zbb$ vertex

Let us consider the one-family model which was introduced in Ref.4. The gauge group is $SU(N_{TC}+1)_{ETC} \times SU(3)_C \times SU(2)_L \times U(1)_Y$, and the fermion contents are

$$\begin{pmatrix} \begin{pmatrix} U^1 & \cdots & U^{N_{TC}} & t \end{pmatrix}_L \\ \begin{pmatrix} D^1 & \cdots & D^{N_{TC}} & b \end{pmatrix}_L \end{pmatrix} \sim (N_{TC}+1,\ 3,\ 2,\ 1/6), \tag{2·1}$$

$$\begin{pmatrix} U^1 & \cdots & U^{N_{TC}} & t \end{pmatrix}_R \sim (N_{TC}+1,\ 3,\ 1,\ 2/3), \tag{2·2}$$

$$\begin{pmatrix} D^1 & \cdots & D^{N_{TC}} & b \end{pmatrix}_R \quad \sim \quad (N_{TC}+1,\ 3,\ 1,\ -1/3). \tag{2·3}$$

The lepton sector of the third generation and the first and second generations are omitted from our discussion for simplicity. By the breaking of the ETC gauge group $SU(N_{TC}+1)_{ETC}$ down to the technicolor gauge group $SU(N_{TC})$, two kinds of massive gauge bosons are generated: massive technicolored sideways gauge boson which mediates the transition between the ordinary quarks and the techni-quarks, and massive diagonal gauge boson which flavor-diagonally couples with both the ordinary quarks and the techni-quarks.

In this naive model the masses of the top and bottom quark are degenerate for isospin invariant techni-quark condensates, $\langle \bar{U}U \rangle = \langle \bar{D}D \rangle$, because of the common mass and coupling of the sideways boson for each quarks. Instead of considering an explicit realistic ETC model that realizes $m_t \gg m_b$, we effectively introduce different ETC gauge boson couplings for the two right-handed multiplets, while keeping the technicolor interaction vector-like.

More explicitly, we assign the sideways coupling $g_t\xi_t$ to the left-handed multiplet, $g_t/\xi_t$ to the right-handed multiplet with the top quark, and $g_t/\xi_b$ to the right-handed multiplet with the bottom quark. The mass of the top quark is then given by

$$m_t \simeq \frac{g_t^2}{M_S^2} 4\pi F_\pi^3 \sqrt{\frac{N_C}{N_{TC}}}, \tag{2·4}$$

where $N_C = 3$. The scale $M_S$ is the mass of the sideways boson and the relation $\langle \bar{U}U \rangle \simeq 4\pi F_\pi^3 \sqrt{N_C/N_{TC}}$ (from the naive dimensional analysis[11]) and the leading $1/N$ behavior) is used. The value of the decay constant $F_\pi$ in this model with four weak doublets is $F_\pi = \sqrt{v_{SM}^2/4} \simeq 125$ GeV. Large top quark mass indicates large value of $g_t$ or small value of $M_S$. The bottom quark mass is given by $m_b = \frac{\xi_t}{\xi_b} m_t$. Since we are assuming that the sideways effect can be treated perturbatively, the condition

$$\frac{(g_t\xi_t)^2}{4\pi} < 1 \quad \text{and} \quad \frac{(g_t/\xi_t)^2}{4\pi} < 1, \tag{2·5}$$

is required. The possible range of $\xi_t$ is restricted by this condition.

The couplings of the diagonal ETC boson are fixed by the sideways couplings. For techni-fermions, we obtain the diagonal couplings by multiplying the factor $-\frac{1}{N_{TC}}\sqrt{\frac{N_{TC}}{N_{TC}+1}}$ to their sideways couplings. For quarks, we obtain them by multiplying the factor $\sqrt{\frac{N_{TC}}{N_{TC}+1}}$ to their sideways couplings. These factors are determined by the normalization and traceless property of the diagonal generator of the ETC gauge group. The diagonal interaction is also chiral in the same way as the sideways interaction.

The effect of the sideways and diagonal ETC boson exchange is described by the effective four-fermion interactions at low energy (weak boson mass scale). The techni-fermion currents in the effective four-fermion interactions can be replaced by the corresponding currents of the low energy effective Lagrangian in which the chiral symmetry of the techni-fermion is non-linearly realized. The electroweak symmetry $SU(2)_L \times U(1)_Y$ is gauged in the effective Lagrangian. Only the currents which couple with the weak gauge bosons remain non-zero in the unitary gauge of the effective Lagrangian, because the effective Lagrangian is described only by the would-be Nambu-Goldstone bosons. Those currents are described by the weak bosons. We carry out the following replacements:

$$\bar{Q}_L \frac{\tau^3}{2} \gamma_\mu Q_L \to \frac{1}{4} F_\pi^2 N_C g_Z Z_\mu, \tag{2·6}$$

$$\bar{Q}_R \frac{\tau^3}{2} \gamma_\mu Q_R \to -\frac{1}{4} F_\pi^2 N_C g_Z Z_\mu, \tag{2·7}$$

where $g_Z \equiv \sqrt{g^2 + g'^2}$ with the gauge couplings $g$ and $g'$ of the $SU(2)_L$ and $U(1)_Y$, respectively. Then, we can extract the $Zbb$ vertex correction from the effective four-fermion interaction[3].

The sideways contribution is obtained as[3]

$$(\delta g_L^b)_{\text{sideways}} = \frac{1}{4} \frac{g_t^2}{M_S^2} \xi_t^2 F_\pi^2 g_Z \simeq \frac{1}{4} \xi_t^2 \frac{m_t}{4\pi F_\pi} \sqrt{\frac{N_{TC}}{N_C}} g_Z, \tag{2·8}$$

where Eq.(2·4) is used in the second equality. In the tree level of the standard model, $g_L^b = g_Z(-\frac{1}{2} + \frac{1}{3}s^2)$ with $s \equiv \sin\theta_W = g'/g_Z$. The diagonal contribution is obtained as[5]

$$\begin{aligned}(\delta g_L^b)_{\text{diagonal}} &= -\frac{1}{2} \frac{g_t^2}{M_D^2} F_\pi^2 \frac{N_C}{N_{TC}+1} g_Z \\ &\simeq -\frac{1}{2} \cdot \frac{N_C}{N_{TC}+1} \cdot \frac{m_t}{4\pi F_\pi} \sqrt{\frac{N_{TC}}{N_C}} g_Z,\end{aligned} \tag{2·9}$$

where we neglect the small contribution which is proportional to $\xi_t/\xi_b$ and assume $M_D \simeq M_S$. Therefore, the total correction due to the ETC bosons are obtained as [*)]

$$(\delta g_L^b)_{ETC} = \left( \xi_t^2 - \frac{2N_C}{N_{TC}+1} \right) \frac{m_t}{16\pi F_\pi} \sqrt{\frac{N_{TC}}{N_C}} g_Z. \tag{2·10}$$

To analyze the $Zbb$ vertex, it is convenient to introduce the form factor $\bar{\delta}_b(q^2)$ in terms of which the $Zb_Lb_L$ vertex function is expressed as[2]

$$\Gamma_L^{Zbb}(q^2) = -\hat{g}_Z \left\{ -\frac{1}{2} \left[ 1 + \bar{\delta}_b(q^2) \right] + \frac{1}{3} \hat{s}^2 \left[ 1 + \Gamma_1^{b_L}(q^2) \right] \right\}. \tag{2·11}$$

---

[*)] The overall normalization of the correction becomes a little smaller, if the technicolor dynamics realizes large anomalous dimension of the techni-fermion mass operator to suppress the flavor-changing neutral current[12, 13].

The hatted quantities, $\hat{g}_Z$ and $\hat{s}$, are the $\overline{MS}$ couplings, and the form factor $\Gamma_1^{b_L}(q^2)$ is small in the standard model. The correction due to the ETC bosons is translated as

$$\bar{\delta}_b(m_Z^2)_{ETC} = \left(\frac{2N_C}{N_{TC}+1} - \xi_t^2\right) \frac{m_t}{8\pi F_\pi}\sqrt{\frac{N_{TC}}{N_C}}. \tag{2·12}$$

The correction within the standard model has been estimated. The one-loop correction[2] and the two-loop correction of $\mathcal{O}(\alpha_s m_t^2)$[14] is parameterized as

$$\bar{\delta}_b(m_Z^2)_{SM} = -0.0099 - 0.0009\frac{m_t - 175\text{GeV}}{10\text{GeV}} \tag{2·13}$$

for $\alpha_s = 0.11 \sim 0.12$ and $m_t = (160 \sim 190)$GeV. We can neglect the $\mathcal{O}(m_t^4)$ two-loop contribution which is about one order smaller than the $\mathcal{O}(\alpha_s m_t^2)$ contribution.

From the measurement of $R_b$, we can obtain the constraint on $\bar{\delta}_b(m_Z^2)$ without uncertainty of $\alpha_s$ and universal oblique correction[15]:

$$\bar{\delta}_b(m_Z^2) = 0.0011 \pm 0.0051, \tag{2·14}$$

which is about 2-$\sigma$ away from the standard-model prediction Eq.(2·13). If this deviation is due to the new physics, the experimental constraint on the new contribution to the $Zb_Lb_L$ vertex is

$$\bar{\delta}_b(m_Z^2)_{new} = 0.0110 \pm 0.0051 + 0.0009\frac{m_t - 175\text{GeV}}{10\text{GeV}}. \tag{2·15}$$

If the ETC contribution Eq.(2·12) dominates the difference Eq.(2·15), we find the following constraint

$$\left(\frac{2N_C}{N_{TC}+1} - \xi_t^2\right)\sqrt{\frac{N_{TC}}{N_C}} = 0.20 \pm 0.09 + 0.005\frac{m_t - 175\text{GeV}}{10\text{GeV}}, \tag{2·16}$$

where we take $F_\pi = 125$ GeV.

The possible value of $N_{TC}$ and the range of $\xi_t^2$ are constrained also by the mass formula of the top quark, Eq.(2·4), and the perturbative condition, Eq.(2·5). If we take the ETC scale $M_S \simeq M_D = 1$ TeV, the values $N_{TC} = 2, 3, \cdots, 8$ are possible. The minimal and maximal values of $\xi_t^2$ allowed by the perturbative condition of Eq.(2·5) for $M_S = 1$ TeV and the experimental constraint from Eq.(2·16) for $m_t = 175$ GeV are shown in Table I for several $N_{TC}$ values. We find that the condition of Eq.(2·16) can be naturally satisfied in the range $2 \leq N_{TC} \leq 5$. It is worth noting here that the cancelation between the sideways and the diagonal contributions naturally explain the LEP result for reasonable range of $N_{TC}$ and $\xi_t^2 = \mathcal{O}(1)$.

Next we discuss the value of $\alpha_s(m_Z)$ which is extracted from the global fit of the LEP data. It is worth noting that the three accurately measured quantities, $\Gamma_Z$, $R_l = \Gamma_h/\Gamma_l$, and $\sigma_h^0$, determine just one combination of $\alpha_s(m_Z)$ and $\bar{\delta}_b(m_Z^2)$, $\alpha_s' = \alpha_s(m_Z) + 1.6\bar{\delta}_b(m_Z^2)$[2], since these observables depend on $\alpha_s$ and the $Zb_Lb_L$ vertex correction only through one quantity, the hadronic width of the $Z$ boson $\Gamma_h$. Therefore, the significant new physics contribution to the $Zb_Lb_L$ vertex correction affects the $\alpha_s(m_Z)$ value extracted from the electroweak $Z$ observables[2,16]. Moreover, since the above $Z$ observables depend also on the universal oblique correction parameters $S$ and $T$[9], the $\alpha_s(m_Z)$ value extracted from the $Z$ boson data should necessarily depend on the three parameters $S$, $T$, and $\bar{\delta}_b(m_Z^2)$. The global fit to extract the value of $\alpha_s(m_Z)$ has been performed in Ref.15. In terms of the three charge form factors $\bar{g}_Z^2(m_Z^2)$, $\bar{s}^2(m_Z^2)$, and $\bar{\delta}_b(m_Z^2)$ of Ref.2 *), one finds

$$\begin{aligned}\alpha_s(m_Z) &= 0.1150 \pm 0.0044 \\ &- 0.0032\frac{\bar{g}_Z^2(m_Z^2) - 0.55550}{0.00101} + 0.0015\frac{\bar{s}^2(m_Z^2) - 0.23068}{0.00042} - 0.0042\frac{\bar{\delta}_b(m_Z^2) + 0.0034}{0.0026}\end{aligned} \tag{2·17}$$

where $\bar{g}_Z^2(m_Z^2) = 0.55550 \pm 0.00101$, $\bar{s}^2(m_Z^2) = 0.23068 \pm 0.00042$, and

$$\bar{\delta}_b(m_Z^2) = -0.0034 \pm 0.0026 \tag{2·18}$$

are the best fit values and their 1-$\sigma$ errors. It should be noted that the global constraint of Eq.(2·18) is consistent with the constraint of Eq.(2·14) from the $R_b$ data alone, while it is still more than 2-$\sigma$ away from the standard-model prediction (2·13). The value of (2·18) also can be explained by ETC contribution.

Table I. Possible ranges of $\xi_t^2$ for each $N_{TC}$

| $N_{TC}$ | $(\xi_t^2)_{\rm min}$ | $(\xi_t^2)_{\rm max}$ | $(\xi_t^2)_{\rm exp}$ | $\frac{2N_C}{N_{TC}+1}$ |
|---|---|---|---|---|
| 2 | 0.48 | 2.1 | $1.8 \pm 0.11$ | 2 |
| 3 | 0.59 | 1.7 | $1.3 \pm 0.09$ | 1.5 |
| 4 | 0.68 | 1.5 | $1.0 \pm 0.08$ | 1.2 |
| 5 | 0.76 | 1.3 | $0.85 \pm 0.07$ | 1 |
| 6 | 0.83 | 1.2 | $0.72 \pm 0.06$ | 0.86 |
| 7 | 0.90 | 1.1 | $0.62 \pm 0.06$ | 0.75 |
| 8 | 0.96 | 1.0 | $0.54 \pm 0.06$ | 0.67 |

The value of $\alpha_s(m_Z)$ which is obtained from the global fit (2·17), $\alpha_s(m_Z) = 0.1150 \pm 0.0044$, is highly consistent with the average value of the results given by the Lattice-QCD analyses of the bottomonium system[7], $\alpha_s(m_Z) = 0.115 \pm 0.002$, and also with the global average value by the Particle Data Group[8], $\alpha_s(m_Z^2) = 0.117 \pm 0.005$.

*) For a given set of $m_t$ and $m_H$, the values of $\bar{g}_Z^2(m_Z^2)$ and $\bar{s}^2(m_Z^2)$ are determined in terms of the values of $S$ and $T$ parameters.

## §3. Diagonal ETC boson and $T$ parameter

We showed in the previous section that the diagonal ETC boson plays an important role to explain the experimental data on the $Zbb$ vertex. But Yoshikawa has pointed out that the diagonal ETC boson gives too large contribution to the $T$ parameter[10]. In the following we estimate the contribution following his method.

The effect of the diagonal ETC boson exchange is described by the effective four-fermion interaction at low energy, as explained before. The four-fermion interaction which includes only the techni-quarks are contained in the four-fermion interaction generated by the diagonal ETC boson exchange. Basing on the factorization hypothesis and the vacuum insertion approximation, we replace the techni-fermion currents by the corresponding currents of the low energy effective Lagrangian, and obtain the correction to the masses of $Z$ and $W$ bosons, $\delta m_Z^2$ and $\delta m_W^2$, respectively. We carry out the following replacements in addition to the replacements of Eqs.(2·6) and (2·7):

$$\bar{Q}_L \frac{\tau^1}{2} \gamma_\mu Q_L \rightarrow \frac{1}{4} F_\pi^2 N_C g W_\mu^1, \qquad (3\cdot19)$$

$$\bar{Q}_R \frac{\tau^1}{2} \gamma_\mu Q_R \rightarrow -\frac{1}{4} F_\pi^2 N_C g W_\mu^1. \qquad (3\cdot20)$$

We obtain the contribution of the diagonal ETC boson to the $T$ parameter as

$$T_{diag} = \frac{1}{2\alpha}\left(\frac{\delta m_W^2}{m_W^2} - \frac{\delta m_Z^2}{m_Z^2}\right) = \frac{1}{16 c^2 s^2} \frac{m_t F_\pi}{m_Z^2} \frac{N_C + 1}{N_{TC} + 1} \sqrt{\frac{N_C}{N_{TC}}} \left(1 - \frac{m_b}{m_t}\right)^2 \frac{1}{\xi_t^2}, \qquad (3\cdot21)$$

where $c^2 = 1 - s^2$ and we use the relation $\xi_t/\xi_b = m_b/m_t$. Note that this contribution originates in the large weak isospin violation $m_t \gg m_b$. The $\xi_t^2$ dependence of $T_{diag}$ is shown in Fig.1. The parameter $\xi_t^2$ must be large for small $T$ parameter, since the large $\xi_t^2$ means the small ETC couplings for right-handed multiplets, and the small ETC couplings for right-handed multiplets means the small weak isospin violation. (Remember that we assigned sideways coupling $g_t/\xi_t$ to the right-handed multiplet with the top quark and $g_t/\xi_b$ to the right-handed multiplet with the bottom quark, and the ratio of $\xi_t$ and $\xi_b$ is fixed by $\xi_t/\xi_b = m_b/m_t$.) But the large value of $\xi_t^2$ results small or negative contribution to the form factor $\bar{\delta}_b(m_Z^2)$ (see Eq.(2·12)), and the experimental value of $R_b$ is not explained. By comparing Table I and Fig.1, we see that there is no possible region of $\xi_t^2$ in which both the constraints from $R_b$ and $T$ parameter are simultaneously satisfied, except for the very narrow region $\xi_t^2 \simeq 0.9$ when $N_{TC} = 5$. Therefore, some mechanisms which generate the negative contribution to the $T$ parameter are expected to explain the experimental value of $R_b$ by the correction due to the ETC gauge bosons.

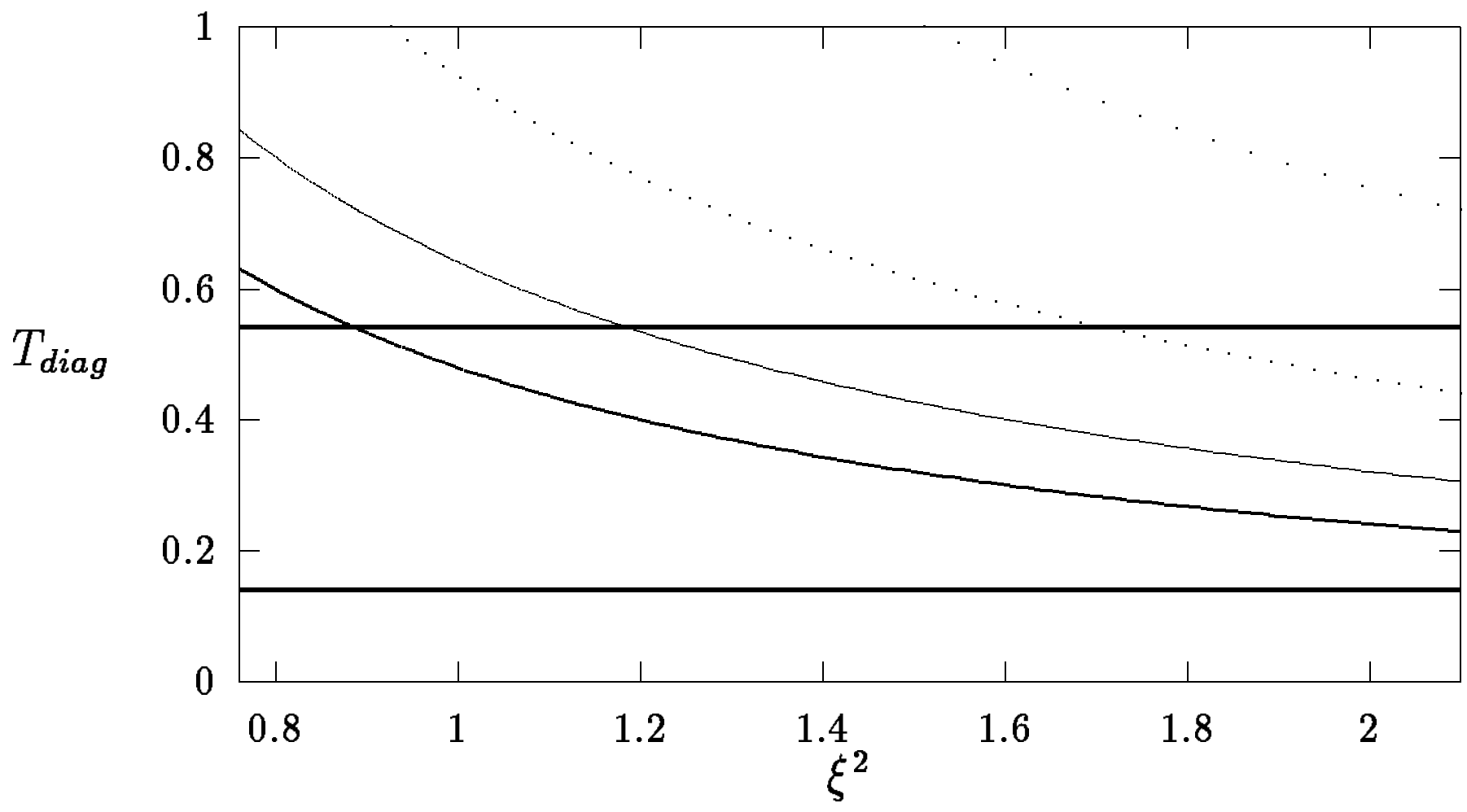

Fig. 1. The $\xi_t^2$ dependence of $T_{diag}$. The region between the two horizontal lines is the experimentally allowed region. The curves are corresponding to the case $N_{TC}$ = 2, 3, 4, and 5 from up to down, respectively.

## §4. Conclusion

We showed that the deviation of the LEP result on $R_b$ from the standard-model prediction can be naturally explained in the ETC theory. Since the diagonal and sideways contributions to the $Zbb$ vertex are opposite in sign and individually larger than the contribution of the standard model, the model can naturally explain the 2-$\sigma$ discrepancy from the standard-model prediction for the reasonable values of $N_{TC}$ and the ETC couplings. The value of $\alpha_s(m_Z)$ which is extracted from the $Z$ boson data becomes small by considering the correction from ETC. The value is consistent with the recent Lattice-QCD estimate and the global average value by the Particle Data Group, but is somewhat smaller than that extracted from jet analysis[8)].

We also showed that this solution to explain the deviation of the LEP result on $R_b$ has a problem of large $T$ parameter which has been pointed out by Yoshikawa[10)]. Because of the large weak isospin violation in the ETC coupling of the right-handed multiplets, the diagonal ETC boson generate too large positive contribution to the $T$ parameter in comparison with the experimental constraint. Therefore, if the value of $R_b$ really deviates form the standard-model prediction by the ETC effect, we should have some mechanisms which generate negative contribution to the $T$ parameter. Recently, a one-family technicolor model which is consistent with the experimental constraint on all the $S$, $T$, and $U$ parameters is proposed by Yanagida and author[17)]. The model has the possibility to generate rather large negative contribution to the $T$ parameter by virtue of the Majorana mass of the right-handed techni-neutrino.

## Acknowledgements

We are grateful to T. Yoshikawa for the comments on the contribution of the diagonal ETC boson to the $T$ parameter.